\documentclass[letterpaper,twocolumn,10pt]{article}
\usepackage{epsfig,endnotes}

\usepackage{microtype}
\newcommand{\rtitle}{Similarity-based matching meets Malware Diversity}
\usepackage{amsmath}
\usepackage{amssymb}
\usepackage{amsthm}
\usepackage[small,compact]{titlesec}
\usepackage{todonotes}
\usepackage{cite}

\usepackage[
 pdftitle={\rtitle},
 pdfsubject={\rtitle},
 pdfauthor={},
 pdfkeywords={}
]{hyperref}
\usepackage{tikz}
\usepackage{pgfplots}
\begin{document}
%\linespread{0.97}

\def\sectionautorefname{Section}
\def\subsectionautorefname{Section}
\def\subsubsectionautorefname{Section}
\def\figureautorefname{Figure}
\def\lstlistingautorefname{Listing}

%don't want date printed
\date{}

%make title bold and 14 pt font (Latex default is non-bold, 16 pt)
%\title{\vspace{-0.5in}\Large \bf \rtitle}
\title{Similarity-based matching meets Malware Diversity}

%for single author (just remove % characters)
\author{
%  {\rm Author names removed for double blind review}
%\normalsize
Mathias Payer$^1$, Stephen Crane$^2$, Per Larsen$^2$, \\
Stefan Brunthaler$^2$, Richard Wartell$^3$, Michael Franz$^2$ \\
$^1$ UC Berkeley and Purdue University, $^2$ UC Irvine, $^3$ Mandiant Corporation
} % end author
\maketitle

% Use the following at camera-ready time to suppress page numbers.
% Comment it out when you first submit the paper for review.
%\thispagestyle{empty}

%\setlength{\droptitle}{-5pt}

%\vspace{-1.3in}
%%%%%%%%%%%%%%%%%%%%%%
%\begin{abstract}
%%%%%%%%%%%%%%%%%%%%%%

\emph{Abstract:} Similarity metrics, e.g., signatures as used by anti-virus products, are the
dominant technique to detect if a given binary is malware. The underlying
assumption of this approach is that all instances of a malware (or even malware
family) will be similar to each other.

Software diversification is a probabilistic technique that uses code and data 
randomization and expressiveness in the target instruction set to generate large
amounts of functionally equivalent but different binaries. \emph{Malware
diversity} builds on software diversity and ensures that any two diversified
instances of the same malware have low similarity (according to a set of
similarity metrics). An LLVM-based prototype implementation diversifies both
code and data of binaries and our evaluation shows that signatures based on
similarity only match one or few instances in a pool of diversified binaries 
generated from the same source code.

%Malware diversity has different constraints compared to defensive software
%diversification: we (i) diversify data alongside with code, instead of only
%diversifying code; (ii) maximize the (byte-wise) differences, minimize the
%largest common subsequence of shared code/data, and randomize the control flow;
%and (iii) ensure that the diversification engine is portable to dynamically
%diversify new binaries on-the-fly. The evaluation of our diversification engine
%shows that signature based matching for these binaries is \emph{severely
%  limited}. Signatures no longer match all instances of a malware version but
%only a single instance or few diversified instances.

%\end{abstract}

%%%%%%%%%%%%%%%%%%%%%%
\section{Introduction}
%%%%%%%%%%%%%%%%%%%%%%

The malware (malicious software) landscape is constantly evolving.  There are no
longer tens of thousands of different malware threats that are currently active
but only few different malware families that often share common
source code. Current malware detection engines (malware scanners and anti-virus
engines) use a combination of signatures, partial matching, regular expressions,
and heuristics to classify binaries as either malicious or benign. Malware
therefore faces a detection problem on current systems due to shared source code
and a low number of currently active malware families. 

Current malware addresses the detection problem using packers~\cite{aspack,
themida} (small pieces of code that obfuscate the actual malware code from
analysis), semi-automatically generating new binaries every couple of hours.
Even Symantec, one of the top anti-virus companies, declares that signature-based
similarity metrics are no longer effective against top
threats~\cite{symantec14av}.  Packers usually work on binaries, however, and are therefore
limited in the expressiveness of the changes due to missing high level
information.

Software diversity~\cite{cohen93csec, collberg97, forrest97hotos, cox06usenix,
kisserli07cobassa, Lin09DIMVA, franz10npsw, multicompiler, Hiser12SP,
Pappas12SP, Giuffrida12SEC} on the other hand uses a compiler to produce 
functionally equivalent binaries that differ substantially at the
implementation level. Software diversity provides protection with quantifiable 
probability from software exploits that rely on a known data or code layout.
Software diversity is also used to thwart reverse-engineering and tampering; 
up until this point, software diversity has been a defensive capability.

\emph{Malware diversity}~\cite{payer14syscan} tailors software diversification
to the needs of malware authors. Malware diversification ensures that two
diversified binaries share neither large amounts of data nor common instruction
sequences. Regular software diversification (i) typically diversifies code
regions while data regions remain constant, (ii) avoids performance degrading
changes, and (iii) is deterministic, i.e., the diversification is reproducible.
Malware diversification shifts these design decisions: the diversification
engine diversifies both code and data (otherwise signatures could match data),
maximizes the (byte-wise and structural) differences, and minimizes the
largest common subsequence of shared code/data.  Reproducibility is only helpful
to report and troubleshoot bugs in deployed software; this is not of concern to
malware authors who can only reliably test malware prior to launching it.

Obfuscation~\cite{collberg97} is closely related to software diversity as both 
techniques rely on randomizing transformations. Obfuscating transformations
protect binaries against reverse engineering by hiding the implemented algorithm.  
Malware diversity is complementary to obfuscation as it modifies the computation 
of every single instance but debugging and reverse engineering of individual 
instances are not affected.

Our malware diversification engine extends the LLVM-based
multicompiler~\cite{multicompiler, payer14syscan} and changes the code of the
application by replacing instructions, instruction reordering, garbage
insertion, and several types of control-flow randomization like reordering basic
blocks. To change the data of an application, the malware diversification engine
uses different data encodings.
%\newpage

The contributions of this paper are as follows:
\begin{enumerate}
  \item a description of malware diversity, a technique that extends
    software diversification to create malware instances with low similarity to
		each other;
  \item a detailed evaluation of a prototype implementation demonstrating
		the effectiveness of malware diversification along several similarity
    metrics.
\end{enumerate}

%%%%%%%%%%%%%%%%%%%%%%%%%%%%%%%%%%%%%%%%%%%%%%%%%
\section{Background and related work}
%%%%%%%%%%%%%%%%%%%%%%%%%%%%%%%%%%%%%%%%%%%%%%%%%

Malware diversity extends software diversity by combining code and
data diversity (to obfuscate data regions along with the code regions).  Malware
diversification is effective when malware detection mechanisms fail to identify
two diversified binaries using the same signature.  Put differently, we can
measure the effectiveness by finding common features that are present in both
diversified malware samples (according to some similarity metric).

\subsection{Malware detection and evasion}
%%%%%%%%%%%%%%%%%%%%%%%%%%%%%%%%%%%%%%%%%%
\label{sec:malwarematching}

Current malware scanners combine different matching techniques to detect
malicious code. Most systems use a combination of different matching
techniques like hash based matching (using a cryptographic hash for the entire
binary or individual sections of the binary), sequence based matching (if two
binaries share a common sequence they are considered equal), expression based
matching (if the regular expression matches both binaries they are considered
equal), and heuristics based matching (if a binary matches a given heuristic it
is considered malicious). This list is based on ClamAV~\cite{clamav}, a
well-known open-source malware scanner; other scanners rely on similar
techniques.  According to Huang and Tsai~\cite{huang10icc} the average matched
pattern length is longer than 25 bytes and only 43 out of more than
83,000 signatures are shorter than 10 bytes. The likelihood of false positives
decreases with increased matched pattern length.

Several new approaches for both malware detection and malware matching
have been proposed to address the limitations of the existing
signature-based techniques. Approaches like malware
normalization~\cite{christodorescu05malwarenormalization} normalize a
binary to a common form but are limited to predefined obfuscation
patterns and cannot undo high-level transformations like register
reallocation.

Heuristics based malware detection tries to match the behavior of a binary to a
specific sequence of actions (such as system or library calls) when executed in
a sandbox. Malware uses subtle differences between the sandbox and a real
system~\cite{garfinkel07hotos, chen08dsn} to detect virtualized
platforms~\cite{ferrie06symantec, paleari09woot, raffetseder07isc,
rutkowska04blog} and stops execution. Approaches that detect sandbox
evasion~~\cite{balzarotti10ndss, ferrie06symantec, johnson11sp,
kang09vmsec,kolbitsch+11,lau10cv, lindorfer11raid} are useful tools for analysts
but usually too heavyweight to be used on a consumer's machine.

%Malware diversification assumes that the diversified malware uses some start up
%code to evade detection in virtualized environments. With malware diversity, the
%red pill code that detects the presence of virtualization or sandboxing is
%diversified like all other code (and therefore protected from signature based
%matching). The sandbox detection code protects the malware from heuristics based
%detection.

\subsection{Packers and binary polymorphism}\label{sec:packers}

A packer~\cite{kang07worm, martignoni07acsac, oberheide09woot, royal06acsac,
perdisci08patternrec} (or crypter) is an application that obfuscates a malicious
application with the intention to hide it from malware scanners or to make
debugging and reverse engineering harder. Packers are historically based on
encryption but moved to oligomoprhic, polymorphic, and metamorphic
transformations~\cite{szor.ferrie+01, okane.etal+11}. Botnet operators can
strengthen polymorphic transformations by randomizing them on a per-machine
basis, for example by using \texttt{perl} scripts to use non-standard
transformation algorithms~\cite{botnetop12}.

Since packers are complex to construct, many malware authors reuse existing
solutions.  Attackers are increasingly shifting to less common packers,
customized packers, and obfuscating packers~\cite{pandasecurity}.  However,
anti-virus scanners can still detect packed binaries due to their special
characteristics.  These include ``weird'' section names, sections with high
Shannon's entropy due to compression, few imported functions, and unusual entry
code~\cite{fsecure, perdisci08patternrec}.  Finally, many anti-virus scanners can even unpack and scan
the payloads of known packers~\cite{kang07worm, martignoni07acsac,
  royal06acsac}, allowing the use of previously discussed detection
techniques.

% Detection approaches based on unpacking~\cite{kang07worm, martignoni07acsac,
% perdisci08patternrec, royal06acsac} detect existing packers and reverse the
% packing process so that they can match the original binary.  Binaries produced
% by advanced packers~\cite{aspack, themida} are not unpackable by any automated
% tools currently available. 

\subsection{Software diversification}\label{sec:software-diversification}
%%%%%%%%%%%%%%%%%%%%%%%%%%%%%%%%%%%%%

Software diversification~\cite{cohen93csec, collberg97, forrest97hotos,
cox06usenix, kisserli07cobassa, Lin09DIMVA, franz10npsw, multicompiler, Hiser12SP,
Pappas12SP, Giuffrida12SEC} is a promising technique. Diversity can be
used to (i) increase the resilience of software against
attacks~\cite{cohen93csec, forrest97hotos, geer03cciar, cox06usenix,
kisserli07cobassa, Lin09DIMVA, franz10npsw, multicompiler, Hiser12SP, Pappas12SP,
Giuffrida12SEC}, to (ii) hide steganographic messages in
binaries~\cite{elkhalil04icics}, and to (iii) protect software against
tampering~\cite{collberg12acsac}. Software diversification constructs
functionally equivalent programs that differ in their code and/or data layout. A
diversification engine uses several (compiler) techniques to randomize the code and
data comprising an application: (i) instruction replacement and reordering, (ii)
variable substitution, (iii) register reordering, (iv) control flow changes, (v)
adding side-effect free instructions, (vi) instruction set
randomization~\cite{barrantes05tiss, kc03ccs, sovarel05ssym, williams09sp},
(vii) instruction stitching~\cite{DBLP:conf/woot/MohanH12}, or (viii) covert
computation~\cite{schrittwieser13ccs}. Larsen et al.~\cite{larsen.etal+14} 
survey the area of automated software diversity in greater detail. % shameless plug :)

%%%%%%%%%%%%%%%%%%%%%%%%%%%%%%%%%
\section{Malware diversification}
%%%%%%%%%%%%%%%%%%%%%%%%%%%%%%%%%

Malware diversification~\cite{payer14syscan} is a form of software
diversification that focuses on avoiding similarity-based detection by malware
scanners. To this end, malware diversity modifies code and data regions of
binaries. As a result, malware analysts are unable to generate a signature that
matches more than few instances of a malware, if any two binaries only share few
instructions at the same offsets, share no common data, and have dissimilar
control flow graphs. A notable difference to other software diversification
techniques is that static data must be diversified alongside code and static
data (otherwise a signature would just match the static data).

Code diversification for malware builds on existing software diversification
mechanisms like instruction replacement, instruction reordering, register
reordering, and changing control flow by splitting and reordering basic blocks,
inlining, outlining, and adding opaque predicates. Malware diversification
configures software diversification to maximize diversity in the generated code
and to minimize similarity between multiple diversified binaries.

Data diversification changes the encoding of static data in the binary. Due to
limited knowledge of the structure of data, malware diversity resorts to a form
of obfuscation~\cite{collberg97} to hide the actual static data. Our malware
diversification engine uses a simple encoding scheme that is applied to static
data during compilation. All static data is encoded with a random key and simple
arithmetic operations (e.g., \texttt{xor}) decode the data at runtime. The
decoding function is diversified along with all other code. 

The data diversification presented here is only a simple technique that can be
strengthened, e.g., by encrypting the data. Also, most programs have
little static data compared to the amount of code. Note that malware diversification 
does not result in binaries with the same special properties that makes it easy for 
anti-virus software to detect packers (cf. Section~\ref{sec:packers})---rather, 
the resulting binaries look like variations of benign programs. % pl: this is the
% strongest argument i can think of why "malware diversity > packers". however,
% reviewers may argue that malware diversity leads to binaries with another set of
% special characteristics.

%\subsection{Implementation}
%%%%%%%%%%%%%%%%%%%%%%%%%%%

We implemented a simple malware diversification engine on top of the existing
multicompiler~\cite{multicompiler, payer14syscan} that extends the
LLVM~\cite{lattner04cgo} compilation framework version 3.4. The compiler is
organized as a sequence of passes which transform the instruction stream. By
modifying the existing compiler transformations and adding new ones, we enable
malware diversification.  Our prototype currently does not use runtime data
structure diversification~\cite{Lin09DIMVA}.

Unlike traditional compiler optimizations that choose whether to transform the
code or not based on program analysis, heuristics, and profile feedback, our diversifying
transformations use a random number generator to ``flip a coin'' at every
opportunity to diversify. We perform the following forms of diversification: (i)
instruction replacement, randomly swapping \texttt{mov} and \texttt{lea}
instructions, (ii) instruction reordering, (iii) register reordering, (iv)
\texttt{nop} and garbage instruction insertion, (v) control flow randomization
and randomizing the layout of basic blocks, and (vi) static data obfuscation.
The resulting compiled binary is stripped to remove all symbol names (variable
names and function names) in the final diversified binary.

Not only are there several other techniques that we could add to our
diversification engine (cf. Section \ref{sec:software-diversification}), the
diversity generated by each of the existing passes could also be increased.
Consequently, we believe that malware writers will not find it difficult to
replicate our approach. The current implementation targets the IA32 ISA but the
concept is portable to any instruction set and operating system combination.

%%%%%%%%%%%%%%%%%%%%
\section{Evaluation}
%%%%%%%%%%%%%%%%%%%%

This section evaluates the prototype implementation of our malware
diversification engine. Using our prototype implementation and a set of
programs we produce 10 diversified instances for each binary and evaluate the
similarity between different binaries according to a set of metrics.
Unfortunately, a lot of Windows malware is compiled using Visual Studio using
the Microsoft C/C++ compiler. These sources are often not compatible to
Clang/LLVM due to Microsoft specific intrinsics in system C++ headers.  All
benchmarks are executed on an Intel Core i3-3770 CPU with 16GB RAM on Debian 7.1
using the Linux 3.6.1 kernel.

We use the following applications: all C/C++ applications of the SPEC CPU2006
benchmarks, kbackdoor (a simple Windows malware), %nmap (a network port scanner),
a simple port scanner, pwdump (a Windows password recovery tool for LM and NTLM
hashes), and mimikatz (a Windows password recovery tool that targets lsass.exe).
We successfully compiled these programs using our diversifying compiler. Some
of these programs are very small: the simple port scanner and {\tt mimikatz} are
below 10KB, {\tt pwdump} and {\tt kbackdoor} are below 75KB. Only nmap can be
considered a ``large'' binary with 3.4MB. The small size makes is harder for the
diversifying compiler to produce a highly diverse population of binaries.

\subsection{Common subsequences}\label{sec:csub}
%%%%%%%%%%%%%%%%%%%%%%%%%%%%%%%%

\begin{figure*}[ht!]
 \begin{center}
   \includegraphics[width=160mm]{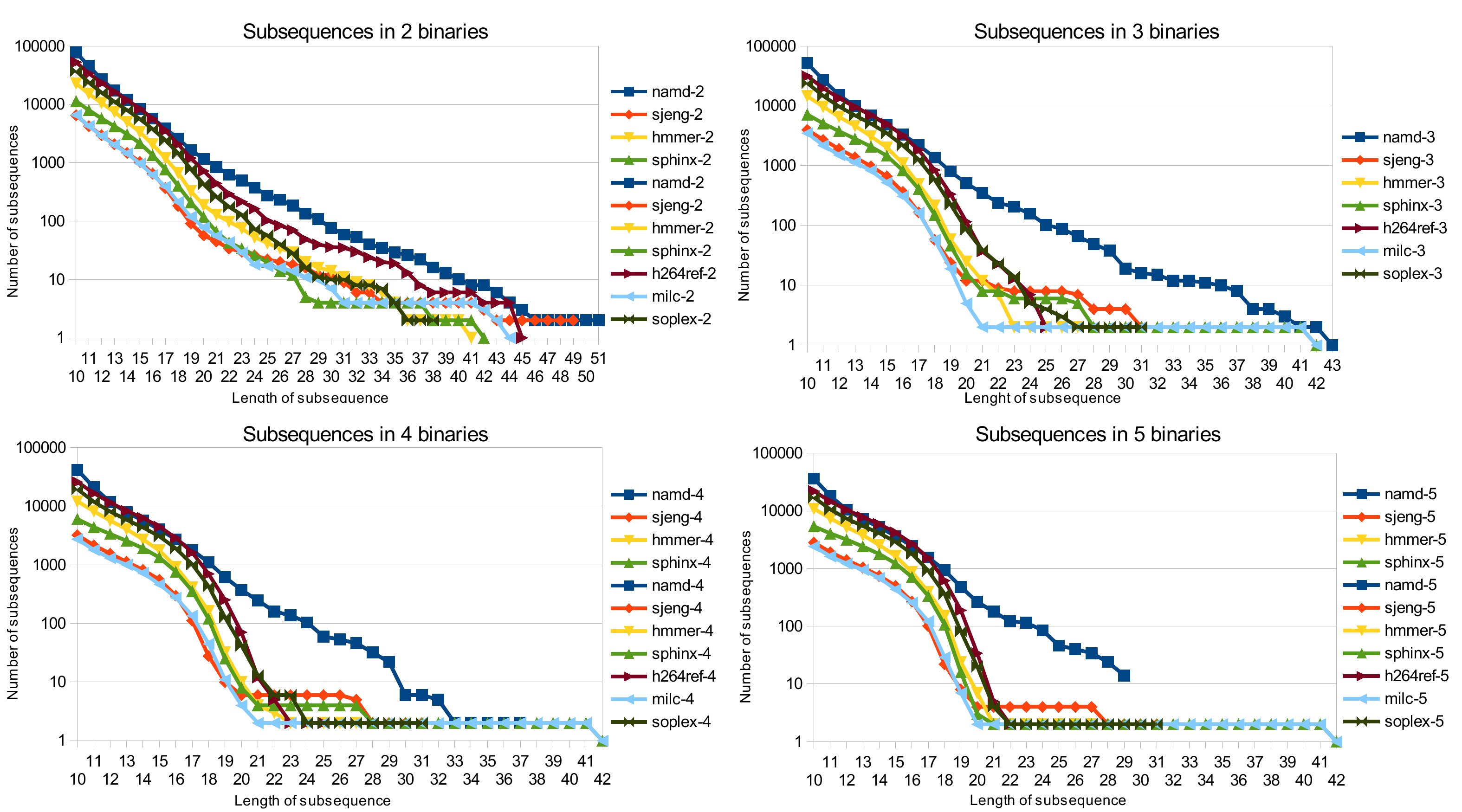}
   \caption{List of common subsequences for SPEC CPU2006 benchmarks (log
     scale).}
   \label{fig:subsequencesspec}
 \end{center}
\end{figure*}

Malware matching relies on signatures: common subsequences that classify the
malware uniquely (see \autoref{sec:malwarematching}). Common subsequences that
are present in many (or all) diversified instances are candidates for
signatures. We search all instances for common subsequences longer than 10
bytes, resulting in an over-approximation of all possible signatures (e.g.,
duplicate code, register spill code, instruction sequences that call library
routines with common parameters, or function prologues and epilogues result in
many shared substrings but are often not usable as signatures).

As indicated by Huang and Tsai~\cite{huang10icc} the average signature is longer
than 25 bytes; shorter signatures are not unique to the malware and lead to a
high number of false positives. \autoref{fig:subsequencesspec} shows common
substrings for diversified versions of SPEC benchmarks shared among a
set of 2, 3, 4, or 5 diversified versions. The figure highlights two interesting
results: (i) the number of shared subsequences of a given length for a benchmark
is lower the more diversified versions are compared (i.e., there are less shared
substrings between three diversified binaries than between two) and (ii) the number
of shared substrings drops logarithmically with increasing length of the
substrings.

The comparison shows that most common sequences are between 10 bytes (10 bytes
is the cutoff length) and 20 bytes of length. Most of these short sequences are
function epilogues and \texttt{nop} sleds to align the next function to a 16
byte address or \texttt{nop} chains before a function prologue. The number of
common subsequences drops drastically with increasing length; only very few
subsequences are longer than 30 bytes.  This comparison shows the effectiveness
of diversification to counter common subsequences in binaries: two different
benchmarks can have higher similarity than three diversified binaries.

We manually looked through the identified subsequences and classified them into
one of the following categories: (i) 10 to 15 byte \texttt{nop} sleds to, e.g.,
align functions to 16 byte offsets, (ii) function call sequences, pushing static
arguments or arguments at specific stack offsets, (iii) \texttt{mov} sequences
that load/store memory into registers, e.g., to initialize structures, (iv)
static start code added by the compiler (e.g., the function that executes before
main is called), and (v) potential signatures. We found that there are only few
potential signature candidates and all of them use registers where register
reordering will introduce diversity for larger sets of binaries.

\subsection{Instruction frequencies and n-grams}
%%%%%%%%%%%%%%%%%%%%%%%%%%%%%%%%%%%%%%%%%%%%%%%%

This similarity metric groups either individual instructions or instruction
mnemonics of a binary, removing register information and memory access
information (e.g., \texttt{mov \%eax, \%ebx} and \texttt{mov \%ecx, \%eax} share
the same instruction mnemonic).  This histogram can be used as a fingerprint of
the malware. We define the similarity measure $S$ between two binaries
$\mathrm{bin}_{1}$ and $\mathrm{bin}_{2}$ as follows:
\[
\mathtt{freq}(\mathrm{mnem}, B) = \frac{\mathrm{mnem}_{\mathrm{total}}(B)}{\mathrm{instrs}_{\mathrm{total}}(B)}
\]
\[
S = 1 - \sum\limits_{\forall i \in \mathrm{instr}} \frac{| \mathtt{freq}(i, \mathrm{bin}_{\mathrm{1}}) -
  \mathtt{freq}(i, \mathrm{bin}_{\mathrm{2}}) |^{2}}{2}
\]

The frequency of one instruction (or mnemonic) is the number of times this
instruction is used in the binary divided by the total number of instructions. A
table of frequencies for each instruction is the histogram of a binary. The
similarity between two binaries is defined as the sum of all absolute squared
differences between each mnemonic frequencies.  Similarity is a natural number
between 1 (every mnemonic occurs an equal number of times in both binaries) and
0 (the two binaries share no instruction mnemonics).

%% \begin{table}[h]
%%   \small
%%   \begin{center}
%%     \begin{tabular}{l|l|l|l|l}
%% &	$bzip2_{1}$	&$ bzip2_{2}$	& $perlb._{1}$	& $perlb._{2}$ \\
%%       \hline
%% $bzip2_{1}$	& 1.000000	& 0.999999	& 0.992362	& 0.994160 \\
%% $bzip2_{2}$	& -	& 1.000000	& 0.992276	& 0.994085 \\
%% $perlb._{1}$	& -	& -	& 1.000000	& 0.999120 \\
%% $perlb._{2}$	&-	& -	& -	& 1.000000 \\
%%     \end{tabular}
%%     \caption{This table shows the similarity between two diversified versions of
%%       each 401.bzip2 and 400.perlbench.}
%%     \label{tbl:similarity}
%%   \end{center}
%% \end{table}

In our experiments with the SPEC CPU2006 benchmarks we found that such simple
fingerprints (both instruction and mnemonic based similarity) are not
significant enough to distinguish diversified binaries of the same program from
other programs with high confidence. For many benchmarks the similarity between
two different benchmarks is as high as the similarity between two diversified
versions of the same benchmark. An interesting observation of this simple
fingerprinting experiment is that different programs with different
functionality have very high similarity. Only few instructions differ overall.

A straight-forward extension of instruction frequencies are n-gram frequencies
where n instructions are bundled together into one class (a 2-gram instruction
sequence would be, e.g., a mov followed by a pop). The simple instruction
frequency defined above represents the 1-gram frequency. We evaluated the
n-gram frequencies for the SPEC benchmarks for $n \in \{2, 3, 4, 5\}$ and found
similar results to $n = 1$: n-gram similarity is not significant enough to match
diversified versions of the same binary. Actually, n-grams offer slightly lower
similarity between diversified versions for $n > 1$ than for $n = 1$.

\subsection{Jaccard similarity}
%%%%%%%%%%%%%%%%%%%%%%%%%%%%%%%

Following the results from the naive malware fingerprinting in the previous
section we refine our similarity metric and use the Jaccard Similarity (JS)
coefficient to compare two binaries.

JS is a statistical metric used to compare the similarity and diversity of 
two sample sets by dividing the size of the intersection of the two sets with 
the size of the union of the two sets:
\[
JS(A, B) = \frac{| A \cap B |}{|A \cup B|}
\]
For each binary we construct a set of instruction frequencies, i.e., we
count for each instruction type how many times it is used in the binary and we
then calculate the JS based on these two sets.

%Consider the following example:
%\begin{alignat*}{2}
%\label{eq:js-example}
%  A        &= \{ &&\mathtt{mov}: 5, \mathtt{imul}: 2, \mathtt{pop}: 23, \mathtt{push}: 25, \\
%  &     &&\mathtt{jmp}: 1 \} \\
%  B        &= \{ &&\mathtt{mov}: 5, \mathtt{imul}: 3, \mathtt{pop}: 25, \mathtt{push}: 25,\\
%  &     &&\mathtt{call}: 1 \} \\
%  A \cap B &= \{ &&\mathtt{mov}: 5, \mathtt{push}: 25 \} \\
%  A \cup B &= \{ &&\mathtt{mov}: 10, \mathtt{imul}: 5, \mathtt{pop}: 48, \mathtt{push}: 50,\\
%  &     &&\mathtt{jmp}: 1, \mathtt{call}: 1 \} \\
%  JS(A, B) &=    &&\frac{2}{6}
%\end{alignat*}

% Two sample sets would be the following:
% $$
% and
% $$.
% For the given example the union of A and
% B is
% $$
% with a size of 6
% and the intersection of A and B is
% $$
% with a size of 2.
% The JS for this example is $$.
The JS effectively highlights differences between sets in terms of instruction
frequences or instuction types. Two binaries that are exactly the same have a JS
of 1 while binaries that share no mnemonics with the same count of instructions
have a JS of 0.

\begin{table*}[ht!]
  \small
  \setlength{\tabcolsep}{1.2pt}
\begin{center}
\begin{tabular}{r|rrrrrrrrrrrrrrrrrrr}
&  1 & 2 & 3 & 4 & 5 & 6 & 7 & 8 & 9 & 10 & 11 & 12 & 13 & 14 & 15 & 16 & 17 &
  18 & 19 \\
\hline
Xalan (1)&\textbf{0.81} & 0.34 & 0.13 & 0.25 & 0.34 & 0.11 & 0.33 & 0.12 & 0.13 & 0.13 &
0.12 & 0.11 & 0.33 & 0.34 & 0.33 & 0.33 & 0.13 & 0.33 & 0.11\\
astar (2)&&\textbf{1.00} & 0.15 & 0.24 & 0.34 & 0.12 & 0.33 & 0.11 & 0.13 & 0.13 & 0.14
& 0.13 & 0.32 & 0.34 & 0.32 & 0.33 & 0.13 & 0.33 & 0.12\\
bzip2 (3)&&&\textbf{0.38} & 0.24 & 0.34 & 0.11 & 0.34 & 0.12 & 0.14 & 0.15 & 0.14 & 0.12
& 0.32 & 0.34 & 0.33 & 0.33 & 0.14 & 0.34 & 0.12\\
dealII (4)&&&&\textbf{0.62} & 0.33 & 0.11 & 0.33 & 0.11 & 0.13 & 0.13 & 0.12 & 0.11 &
0.32 & 0.34 & 0.33 & 0.34 & 0.13 & 0.34 & 0.12\\
gcc (5)&&&&&\textbf{1.00} & 0.12 & 0.33 & 0.11 & 0.14 & 0.13 & 0.12 & 0.12 & 0.32 &
0.33 & 0.33 & 0.33 & 0.13 & 0.34 & 0.11\\
gobmk (6)&&&&&&0.33 & \textbf{0.34} & 0.12 & 0.13 & 0.14 & 0.12 & 0.11 & 0.32 & 0.33 &
0.33 & \textbf{0.34} & 0.13 & \textbf{0.34} & 0.12\\
h264ref (7)&&&&&&&\textbf{1.00} & 0.12 & 0.13 & 0.13 & 0.12 & 0.11 & 0.32 & 0.35 & 0.35 &
0.34 & 0.12 & 0.33 & 0.12\\
hmmer (8)&&&&&&&&0.33 & 0.13 & 0.13 & 0.12 & 0.12 & 0.32 & \textbf{0.34} & 0.32 &
\textbf{0.34} & 0.13 & \textbf{0.34} & 0.12\\
lbm (9)&&&&&&&&&\textbf{0.39} & 0.13 & 0.13 & 0.11 & 0.32 & 0.33 & 0.33 & 0.33 & 0.14
& 0.33 & 0.12\\
libquantum (10)&&&&&&&&&&\textbf{0.38} & 0.13 & 0.12 & 0.32 & 0.34 & 0.33 & 0.33 & 0.14 &
0.33 & 0.12\\
mcf (11)&&&&&&&&&&&\textbf{0.36} & 0.12 & 0.32 & 0.33 & 0.32 & 0.33 & 0.13 & 0.33 &
0.12\\
milc (12)&&&&&&&&&&&&\textbf{0.33} & 0.33 & 0.33 & 0.32 & 0.33 & 0.13 & 0.33 & 0.12\\
namd (13)&&&&&&&&&&&&&\textbf{0.96} & 0.34 & 0.32 & 0.33 & 0.13 & 0.34 & 0.11\\
omnetpp (14)&&&&&&&&&&&&&&\textbf{0.99} & 0.34 & 0.33 & 0.14 & 0.33 & 0.12\\
perlbench (15)&&&&&&&&&&&&&&&\textbf{0.97} & 0.33 & 0.12 & 0.33 & 0.12\\
porvray (16)&&&&&&&&&&&&&&&&\textbf{1.00} & 0.13 & 0.34 & 0.12\\
sjeng (17)&&&&&&&&&&&&&&&&&\textbf{0.37} & 0.33 & 0.12\\
soplex (18)&&&&&&&&&&&&&&&&&&\textbf{0.99} & 0.12\\
sphinx (19)&&&&&&&&&&&&&&&&&&&\textbf{0.34}\\
\end{tabular}
\caption{Jaccard similarity for all C/C++ SPEC benchmarks with themselves and
  with each other. %The X and Y axis each contain different diversified versions
  %of Xalan (1), astar (2), bzip2 (3), dealII (4), gcc (5), gobmk (6), h264ref (7), hmmer
  %(8), lbm (9) libquantum (10), mcf (11), milc (12), namd (13), omnetpp (14),
  %perlbench (15), povray, (16), sjeng (17), soplex (18), sphinx (19). Higher
  %values show higher similarity.
}
\label{tab:similarity}
\end{center}
\end{table*}

\autoref{tab:similarity} shows the JS coefficient measurements of diversified
versions of all C/C++ SPEC CPU2006 benchmarks. Each benchmark is diversified 3
times. If the benchmark is compared with itself then we report the average JS of
all three diversified binaries between each other (e.g., for bin1, bin2, bin3 we
report the average of bin1-bin2, bin2-bin3, and bin1-bin3). If two different
benchmarks are compared then we report the average of all 9 individual JS
between each diversified version of the first and second benchmark.
JS is somewhat effective: some diversified SPEC benchmarks have a higher
similarity with diversified copies of themselves compared to other diversified
benchmarks. Some benchmarks (lbm, libquantum, and milc) have similarity with
themselves of almost 0.5 or higher. These benchmarks are identifiable due to
specific floating point instructions. Other benchmarks that mostly execute
integer instructions are hard to identify as the similarity between different
diversified binaries is close to (or even lower) than other benchmarks.

\begin{table}[t!]
  \small
  \setlength{\tabcolsep}{2.2pt}
\begin{center}
\begin{tabular}{lrrrrr}
&  {\tt kbackdoor} & {\tt mimikatz} & {\tt pwdump} & {\tt nmap} & {\tt sps} \\
\hline
{\tt kbackdoor} &\textbf{0.68} & 0.27 & 0.29 & 0.43 & 0.28\\
{\tt mimikatz} &&\textbf{0.68} & 0.29 & 0.28 & 0.3\\
{\tt pwdump}  &&&\textbf{0.8} & 0.27 & 0.28\\
{\tt nmap} &&&&\textbf{0.71} & 0.28\\
{\tt sps} &&&&&\textbf{0.77}\\
\end{tabular}
\caption{Jaccard similarity for our malware set with themselves and with each
  other. Higher values indicate higher similarity.}
\label{tab:similarity2}
\end{center}
\end{table}

\autoref{tab:similarity2} shows the JS similarity for our set of malware
programs. Due to the small size of these programs and the single purpose the JS
similarity can be used to successfully identify diversified versions. For the
two larger programs (kbackdoor and nmap) the difference between the JS
similarity for diversified binaries and the JS similarity for different binaries
is smaller than for the other benchmarks due to the additional functionality in
these programs.

%Due to the way we generate the sets (i.e., looking at only the instruction
%mnemonic) we discard information about registers, memory addresses, and ordering
%that are used or alternative encodings of the same instruction and the ordering
%of instructions. This observation makes it less of a surprise that JS is more
%effective at classifying diversified binaries than naive fingerprinting as most
%of our current diversification comes from different instruction encodings,
%register allocation, instruction reordering, and basic block modifications
%(splitting, merging, and reordering).

Overall we can conclude that JS is effective in identifying diversified binaries
of some (smaller) benchmarks with particular instruction sequences. On the other
hand JS cannot be used as a general approach to identify any diversified binary
or more complex program. Especially as the sample set of benign binaries grows
larger it will be hard to get enough confidence to identify diversified versions
of a piece of malware relative to all the benign fingerprints.

As a potential countermeasure malware diversity can decrease the JS by adding
additional garbage instructions alongside the diversified
instructions. Attackers can deliberately choose the type and number of garbage
instructions to minimize the JS score as described by De Sutter et al.~\cite{desutter.etal+08}.  Such a scheme distorts the fingerprint of a given binary and lowers the 
utility of the JS metric.  The implementation of this additional distortion 
tactic is left as future work.

\subsection{Graph based similarity}\label{sec:bindiff}
%%%%%%%%%%%%%%%%%%%%%%%%%%%%%%%%%%%

Bindiff is a plugin for the Interactive DisAssembler (IDA) that compares two
binaries and evaluates their structural similarity based on a set of graph based
matching techniques.  Bindiff works by recovering and comparing approximations
of the actual control-flow and call graphs of two diversified binaries.  The
core algorithm is described in the original bindiff paper~\cite{flake04dimva}.
Since bindiff is a commercial tool, now developed by Google, we expect
substantial improvements have been made in the interim.

Due to the graph-based comparision approach, bindiff is unaffected by those of
our
diversification techniques that leave the flow of control unaffected.

%\begin{table}[ht!]
%  \small
%  \setlength{\tabcolsep}{2pt}
%\begin{center}
%\begin{tabular}{lrrrrrr}
%&  bzip2& gcc&  h264ref&  perlbench&  xalan&  netcat \\
%\hline
%normal& 0.649&  0.658&  0.726&  0.692&  0.737&  0.710 \\
%stripped& 0.634&  0.599&  0.688&  0.592&  0.566&  -\\
%\end{tabular}
%\caption{Similarity of diversified versions according to bindiff.}
%\label{tab:similarity3}
%\end{center}
%\end{table}

\begin{table*}[ht!]
  \small
  \setlength{\tabcolsep}{2pt}
%\begin{center}
\centering
\begin{tabular}{r|rrrrrrrrrrrrrrrrrrr}
		&	1	&	2	&	3	&	4	&	5	&	6	&	7	&	8	&	9	&	10	&	11	&	12	&	13	&	14	\\\hline
astar	(1)	&	0.432	&	0.118	&	0.064	&	0.067	&	0.052	&	0.085	&	0.109	&	0.085	&	0.128
&	0.051	&	0.109	&	0.040	&	0.104	&	0.086	\\
bzip2	(2)	&		&	0.310	&	0.046	&	0.050	&	0.042	&	0.112	&	0.099	&	0.085	&	0.095	&
0.048	&	0.084	&	0.049	&	0.107	&	0.074	\\
h264ref	(3)	&		&		&	0.363	&	0.109	&	0.012	&	0.026	&	0.048	&	0.023	&	0.089	&
0.018	&	0.055	&	0.014	&	0.046	&	0.105	\\
hmmer	(4)	&		&		&		&	0.396	&	0.010	&	0.027	&	0.048	&	0.025	&	0.093	&	0.020	&
0.056	&	0.023	&	0.052	&	0.126	\\
kbackdoor	(5)	&		&		&		&		&	0.377	&	0.076	&	0.065	&	0.090	&	0.028	&	0.311	&
0.037	&	0.359	&	0.037	&	0.025	\\
lbm	(6)	&		&		&		&		&		&	0.440	&	0.115	&	0.244	&	0.062	&	0.139	&	0.090	&
0.107	&	0.098	&	0.037	\\
libquantum (7)		&		&		&		&		&		&		&	0.394	&	0.103	&	0.105	&	0.070	&	0.108	&
0.067	&	0.099	&	0.083	\\
mcf	(8)	&		&		&		&		&		&		&		&	0.331	&	0.056	&	0.130	&	0.071	&	0.141	&
0.080	&	0.035	\\
milc	(9)	&		&		&		&		&		&		&		&		&	0.381	&	0.034	&	0.109	&	0.033	&	0.103
&	0.110	\\
mimikatz	(10)	&		&		&		&		&		&		&		&		&		&	0.502	&	0.060	&	0.413	&	0.048
&	0.027	\\
namd	(11)	&		&		&		&		&		&		&		&		&		&		&	0.538	&	0.043	&	0.101	&	0.074
\\
pwdump	(12)	&		&		&		&		&		&		&		&		&		&		&		&	0.482	&	0.046	&	0.022	\\
sjeng		(13) &		&		&		&		&		&		&		&		&		&		&		&		&	0.343	&	0.073	\\
sphinx		(14) &		&		&		&		&		&		&		&		&		&		&		&		&		&	0.402	\\
\end{tabular}
\caption{Similarity of diversified versions according to bindiff.
\label{tab:similarity3}}
%\end{center}
%\vspace{-0.5mm}
\end{table*}

\autoref{tab:similarity3} shows the similarity of a subset of the SPEC
benchmarks and our malware set. When comparing diversified versions of the same
program we use a set of 5 diversified versions and report the lowest similarity
in this set. In our tests, bindiff showed similar similarity across diversified
versions.

In general, bindiff achieves a higher similarity than the Jaccard similarity due
to the combination of multiple different matching algorithms (including some
graph-based matching). On the other hand even bindiff cannot detect very high
similarity (the maximum similarity is 53.8\%). Bindiff-like similarity metrics
can be used to defeat diversity but at the price of additional manual analysis;
it takes several minutes to analyze a pair of binaries by hand using IDA Pro and
bindiff, resulting in the reported low similarity numbers.

We identified function calls to system libraries (e.g. {\tt libc}) as a major 
source of structural information that assisted bindiff in computing similarity.
We plan to substantially extend our set of transformations that add structural 
diversity to binaries. In particular, our experiments indicate that calling
library functions through randomly generated wrapper functions substantially
affects similarities reported by bindiff.  Coppens et al.~\cite{coppens13taco}
studied structural transformations that defeat bindiff to protect software
patches against reverse engineering; we can benefit from this catalog.  An
additional, orthogonal diversification technique weaves a benign application
into the diversified malware and interleaves the instruction stream of both
programs using unused resources in the malicious program to execute additional
superfluous computation.

%%%%%%%%%%%%%%%%%%%%
\section{Discussion}
%%%%%%%%%%%%%%%%%%%%

Diversity reduces the similarity between different instances of a binary enough
to disable direct, similarity-based matching. Malware diversification can use
existing degrees of freedom in the compilation process and the resulting
binaries to adapt to newly proposed counter measures and use additional
diversification (or garbage insertion). This situation will result in yet
another security arms race between attackers and defenders until either the
diversity in the compilation process is exhausted (which is unlikely) or malware
diversity will circumvent all detection mechanisms.

%In a world where malware is routinely diversified malware scanners will
%inevitably evolve to counter the threat.  Based on our evaluation in
%\autoref{sec:csub} and \autoref{sec:malware-fingerprinting}, we expect that
%security vendors will also find that neither signature-based matching nor
%fingerprinting can classify diversified malware with high precision.

%Malware diversification can use existing additional ambiguity in the compilation
%process and the produced binaries to adapt to newly proposed counter measures
%and use additional diversification (or garbage insertion). This situation will
%result in yet another security arms race between attackers and defenders until
%either the diversity in the compilation process is exhausted (which is unlikely)
%or malware diversity will circumvent all detection mechanisms.

%Focusing on the diversification techniques presented in
%\autoref{sec:software-diversification}, we now consider the possibility of
%tailoring the malware scanner to ``reverse'' the transformations prior to
%signature-based matching.

A possible countermeasure is a canonicalization of diversified binaries that
undoes individual diversifications.  In case of \texttt{nop} insertion, we
recognize that stripping all \texttt{nop}s from a binary\footnote{Our discussion
relies on disassembly of malware which often employs anti-disassembly
techniques. We consider this an orthogonal concern and refer to other solutions, 
e.g.~\cite{DBLP:conf/uss/KruegelRVV04}.} does not produce the
original binary before diversification. The reason being that we cannot
distinguish between \texttt{nop}s inserted during diversifications and
\texttt{nop}s inserted for other purposes (e.g., alignment).  However, stripping
out \texttt{nop}s leaves us with a ``canonical'' version of the binary in the
sense that all similar binaries diversified with \texttt{nop} insertion result
in the same canonical binary.  A similar argument holds for instruction
replacement, instruction reordering and register assignment reordering.  In case
of instruction reordering, we can canonicalize the binary by computing
instruction histograms (using only instruction mnemonics in the presence of
register assignment reordering) and then match up basic blocks using the control
flow graph.

In general, we note that the relationship between diversified and canonicalized
binaries is not bijective; that two diversified binaries map to the same
canonical representation makes it likely, but not certain, that they share the
same source code.  Furthermore, reordering more code features (registers,
instructions, basic blocks, functions) increases the likelihood of two, distinct
programs having the same canonical representation.  Even though this makes false
positives a potential concern, we expect that a diversification-aware matching
strategy based on canonicalization is more accurate than any
diversification-oblivious attempt at classifying binaries.  On the other hand,
randomly inserting instruction sequences from benign applications reduces our
ability to compute a canonical version of binaries.  We plan to evaluate this
type of defense as future work.

%%%%%%%%%%%%%%%%%%%%
\section{Conclusion}
%%%%%%%%%%%%%%%%%%%%

Malware detection engines rely on effective and efficient similarity metrics to classify
binaries as malicious or benign.  Malware diversity uses software diversity to
break this assumption and randomly diversifies both code and data of programs.
Similarity-based metrics are no longer effective due to the variations
in the binary layout; our experiments confirm that malware diversity results in
very short common subsequences and breaks other similarity metrics as well. 
The structural metrics used by bindiff, on the other hand, are not efficient
to compute.

Malware diversity enables a new class of malware that generates a virtually
unlimited number of unique malware instances.  Our experiments and the
discussion of protective measurments show that malware diversity is powerful
enough to counter new detection mechanisms by exploiting additional
opportunities for diversification.

\section{Acknowledgments}
We thank Todd M. Jackson and Andrei Homescu for their work on the multicompiler,
Thomas R. Gross, Dan Caselden, Lenx Tao Wei, and Dawn Song for feedback,
discussions, and comments.

\linespread{0.9}
{\footnotesize
\bibliographystyle{acm}
\bibliography{bibliography}}

\end{document}